\documentclass[5p,times]{elsarticle}

\usepackage[misc]{ifsym}
\usepackage{verbatim}
\usepackage{bbding}
\usepackage[nointegrals]{wasysym} 
\usepackage[marginal]{footmisc}
\usepackage{endnotes}
\usepackage{bm}
\usepackage{booktabs}
\usepackage{amssymb}
\usepackage{amsmath}
\usepackage{url}

\begin{document}
\begin{frontmatter}

\title{Deep multi-metric learning for text-independent speaker verification}

\author[add1]{Jiwei Xu}
\author[add1]{Xinggang Wang\corref{corr}}
\ead{xgwang@hust.edu.cn}
\author[add1]{Bin Feng}
\author[add1]{Wenyu Liu}

\address[add1]{School of Electronic Information and Communications, \\ Huazhong University of Science and Technology, Wuhan 430074, China}

\cortext[corr]{Corresponding author}

\begin{abstract}
 Text-independent speaker verification is an important artificial intelligence problem that has a wide spectrum of applications, such as criminal investigation, payment certification, and interest-based customer services. The purpose of text-independent speaker verification is to determine whether two given uncontrolled utterances originate from the same speaker or not. Extracting speech features for each speaker using deep neural networks is a promising direction to explore and a straightforward solution is to train the discriminative feature extraction network by using a metric learning loss function. However, a single loss function often has certain limitations. Thus, we use deep multi-metric learning to address the problem and introduce three different losses for this problem, i.e., triplet loss, n-pair loss and angular loss. The three loss functions work in a cooperative way to train a feature extraction network equipped with Residual connections and squeeze-and-excitation attention. We conduct experiments on the large-scale \texttt{VoxCeleb2} dataset, which contains over a million utterances from over $6,000$ speakers, and the proposed deep neural network obtains an equal error rate of $3.48\%$, which is a very competitive result. Codes for both training and testing and pretrained models are available at \url{https://github.com/GreatJiweix/DmmlTiSV}, which is the first publicly available code repository for large-scale text-independent speaker verification with performance on par with the state-of-the-art systems.
 \end{abstract}


\end{frontmatter}

\section{Introduction}
{SV (speaker verification) is a key technology for intelligent interaction. It can be widely used in financial payment, criminal investigation, national defense and other fields. It is one application in speech recognition that aims to verify a claimed identity based on his/her utterance \cite{zhang2018text}.} This task is a $1:1$ match where one speaker's voice is matched to a particular template. SV can be categorized into text-dependent and text-independent \cite{hansen2015speaker, variani2014deep}. The text-dependent SV system requires the speech to be produced from a fixed or prompted text phrase, while the text-independent SV system operates on unconstrained speech. Therefore, text-independent SV is a more challenging problem, but it is more useful in practical applications.

\begin{figure}[htbp]
\centering
\includegraphics[width=0.4\linewidth]{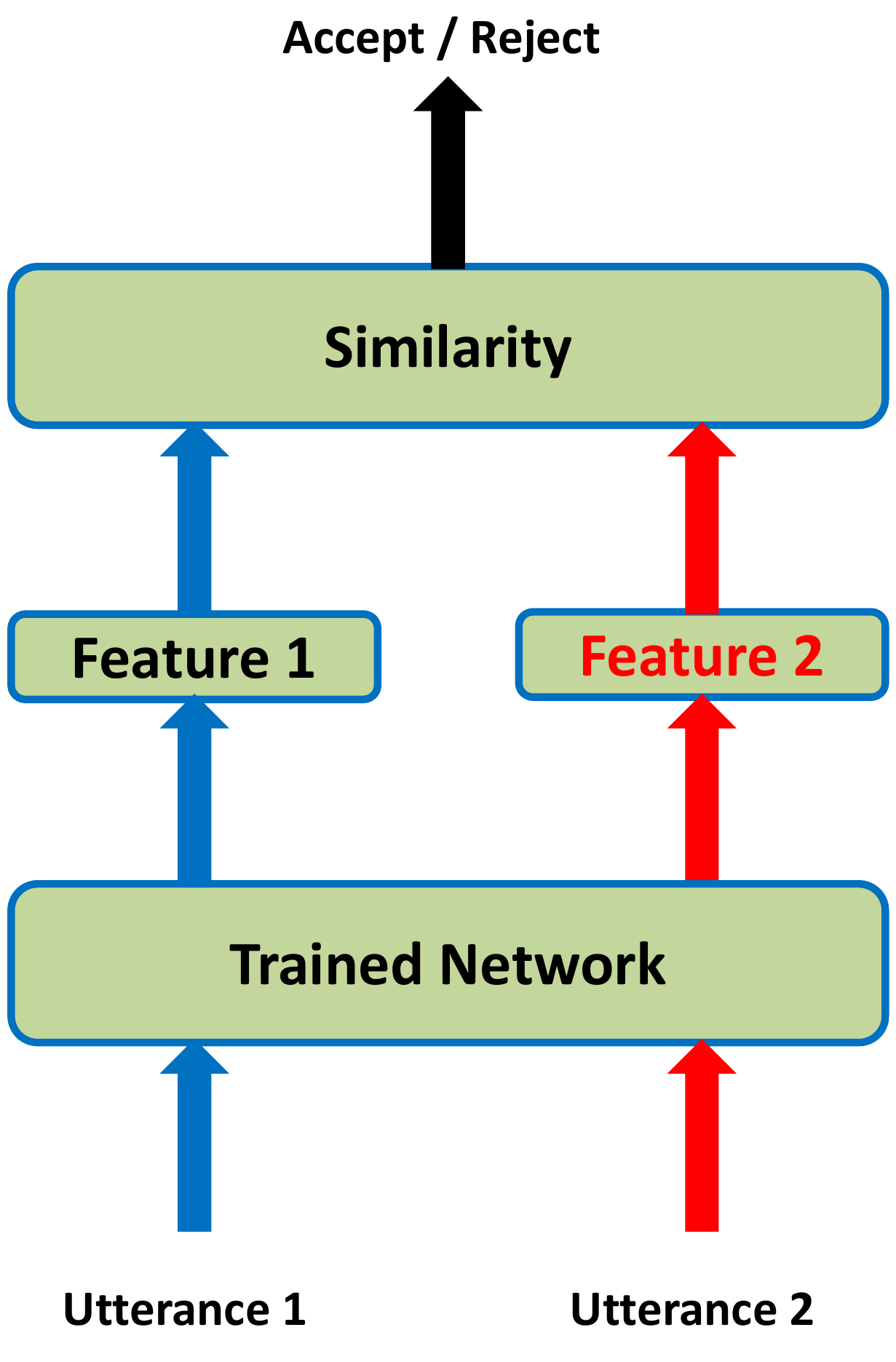}
\caption{Illustration of SV using a deep metric learning network.}
\label{fig:train_test}
\end{figure}

Generally, a deep learning-based SV system contains the training step and testing step \cite{variani2014deep}. In the training step, we use a large collection of utterances to train an SV neural network. The learned deep neural network model is used as a universal feature extractor for any testing speaker. Then in the testing step, two different utterances are separately sent to the learned deep model for feature extraction, and we compute the similarity based on the two feature vectors to perform SV. In the test phase, the false acceptance/rejection rates depend on the predefined threshold \cite{rahman2018attention}. The equal error rate (EER) metric projects the error when the two aforementioned rates are equal. The basic training verification system is shown in Figure~\ref{fig:train_test}.

Before the era of deep learning, traditional SV models made remarkable achievements. For example, the Gaussian mixture model with a universal background model (GMM-UBM) \cite{larcher2014text} uses a sufficiently large speech dataset of several hours from multiple sources. For a D-dimensional feature vector $\bm{x}$, the mixture density used for the likelihood function is defined as
\begin{equation}
P( \bm{x} | \lambda) = \sum_{i=1}^{M}w_{i}p_{i}(\bm{x}).
\end{equation}
The density is a weighted linear combination of $\bm{M}$ unimodal Gaussian densities, $p_{i} (\bm{x})$, each of which is parameterized by a mean $D \times 1$ vector, $\mu$, and a $D \times D$ covariance matrix $\Sigma_{i}$ \cite{reynolds2000speaker}:
\begin{equation}
P_{i}(\bm{x}) = \frac{1}{(2\pi)^{D/2}|\Sigma_{i}|^{1/2}} \exp \left\{ -\frac{1}{2}(\bm{x}-\bm{\mu}_{i})^{'}(\Sigma_{i}^{-1})(\bm{x}-\bm{\mu}_{i})\right\}.
\end{equation}

Furthermore, the mixture weights $w_{i}$ satisfy the constraint $\Sigma_{i=1}^{M}w_{i}=1$. The GMM-UBM system is a straightforward generative approach for an SV task, using a sufficiently large speech data sample of several hours from multiple sources. The UBM is represented as follows:
\begin{equation}
\mathbf{\lambda = (w, \mu, \Sigma)},
\end{equation}
where $\bm{w}$ is the weight of the $i$-th Gaussian component, $\bm{\mu}$ represents the mean and $\bm{\Sigma}$ is the covariance matrix of the $i$-th Gaussian component. Each speaker is represented as a GMM derived by maximum-a-posteriori (MAP) adaptation from UBM. 

Apart from GMM-UBM, i-vector\cite{stafylakis2013text} is another state-of-the-art SV framework. It models the speaker factors and channel factors and converts the utterance of the speaker identity to a low-dimensional embedding representation. The i-vector representation, whose role is to represent an utterance of arbitrary duration by a vector of fixed dimension which we denote by $d$ (in the range of $400$ to $600$), originates from the joint-factor analysis \cite{kenny2008study} method.
The SV systems based on i-vector represent the high-dimensional GMM supervector in a transform vector (TV) space which reduces the supervector into low-dimensional factors. In TV space, the GMM supervector, is projected as
\begin{equation}
\mathbf{M = m + \bm{T}\bm{i}},
\end{equation}
where $\bm{T}$ is a low-rank factor loading matrix, $\bm{m}$ is channel and $\bm{i}$ is the speaker-independent supervector whose prior distribution is assumed to be standard normal:
\begin{equation}
\bm{i} \sim \bm{N(0, I)},
\end{equation}
an in-depth review of these traditional methods is given in \cite{hansen2015speaker}.

Recently deep convolution neural networks (CNN) have witnessed a wide spectrum of applications in computer vision \cite{yin2019fourier, campbell2019computer}, natural language processing (NLP) \cite{ponti2019modeling, ruder2019transfer} and speech recognition \cite{he2019streaming, nassif2019speech}. Motivated by the powerful feature extraction capability and recent successes of deep learning applied to SV, more deep learning-based SV methods have been developed. Recently, \cite{bhattacharya2019deep} has achieved the results of the new state-of-the-art. These works extract bottleneck features from deep neural networks (DNN) that are trained by probabilistic linear discriminant analysis (PLDA) \cite{ju2019probabilistic} or deep metric learning loss (such as contrastive loss \cite{cheng2019modified}, triplet loss \cite{do2019theoretically} or angular softmax Loss \cite{li2019towards}). In \cite{bhattacharya2017deep}, SV systems with DNN are shown to achieve better performance than traditional SV methods. Therefore, a text-independent SV framework based on the deep residual network (ResNet) \cite{he2016deep} is investigated in this study, where a non-fixed length speaker discriminative model is learned from sparse speech features and utilized as a feature representation for SV tasks. {Training ResNet for SV requires metric learning loss functions. In this paper, we find the complementarity among different metrics (e.g., softmax loss, triplet loss, n-pair loss and angular loss) and a multi-metric learning scheme for text-independent SV. In experiments, we demonstrate the superiority of multi-metric training on a large-scale dataset.}

The rest of this paper is organized as follows. In Section~\ref{sec:rw}, previous related work on SV is described. In Section~\ref{sec:mec} we present our method. Section~\ref{sec:exp} shows the experimental results of our SV system. Finally, Section~\ref{sec:con} concludes the paper.

\section{Related Work}
\label{sec:rw}

The traditional SV models, such as GMM-UBM \cite{larcher2014text} and i-vector \cite{stafylakis2013text}, have been the state-of-the-art approaches for a long time. All the above mentioned-methods rely on low dimensional input features extracted by using mel-frequency cepstrum coefficients (MFCC), however, MFCC is known to suffer from performance degradation under real-world noise conditions as demonstrated by \cite{yapanel2002high,hansen2001robust}. Deep Convolutional Neural Networks (DCNN) have proven to be effective to extract intrinsic features from noisy data, thus various speech applications \cite{yadav2018learning, hershey2017cnn, lukic2016speaker} have been proposed based on DCNN. Why is convolution useful for text-independent SV? First, text-independent SV is based on the intrinsic feature of a human utterance, which can be extracted from small fragments in speech. Convolution allows spatial translation invariance and operates on local features. Thus, convolution is more suitable to extract speech features. Second, the convolutional neural network benefits from data augmentation. Thirdly, it is computationally efficient. Ultimately, similar to image classification and face recognition, SV is suitably solved using DCNN.

In recent years, deep learning, especially deep metric learning, has achieved outstanding results in face verification and re-identification problems. The most commonly used metric learning loss function is the triplet loss \cite{do2019theoretically}. The goal of triplet loss is to minimize the distance of the same class pairs and maximize the distance of different class pairs. Although the triplet loss has achieved great results in many tasks, it is restricted to the class imbalance problem and needs a long training time to converge. N-pair loss \cite{chen2019deep} pays more attention to the information of one negative sample in each optimization. To reduce the training burden, each mini-batch selects $N$ pair of examples from $N$ different classes and builds $N$ tuplets to accelerate the model convergence. Different from triplet loss and N-pair loss using the euclidean distance, angular softmax loss \cite{li2019towards} modifies the softmax loss function to learn angularly discriminative embeddings and use a controllable parameter to constraint the intra-speaker variation of the learned embeddings. Based on the above analysis, we believe that the three loss functions exhibit a certain complementarity, and combine them to further optimize the network. Inspired by the recent SENet (squeeze-and-excitation networks) method \cite{zhang2016end, rahman2018attention}, our study on SE block illustrates that different channels of a feature map play different roles in specific objects. The SE block discards the pooling layer and uses $1 \times 1$ convolutional layer to replace the fully-connected layer for learning spatial information. Meanwhile, The SE block is also computationally inexpensive and imposes only a slight increase in model complexity and computational burden.

At present, there are many methods of metric learning that have achieved good results in SV.
\cite{zhang2018text} introduced triplet loss \cite{schroff2015facenet} into SV and achieved very competitive results. \cite{huang2018angular} used angular softmax \cite{li2018angular} achieved an obvious performance improvement compared with other methods in the SV task. The combination of center loss \cite{wen2016discriminative} and softmax loss has also been shown to provide good results in the SV task \cite{li2018deep, yadav2018learning}. Regarding the effectiveness of angular softmax loss, center loss and triplet loss for face verification, it is worth exploring their power in the task of SV, which has never been studied before.

\begin{figure*}[htp]
\centering
\includegraphics[width=0.8\linewidth]{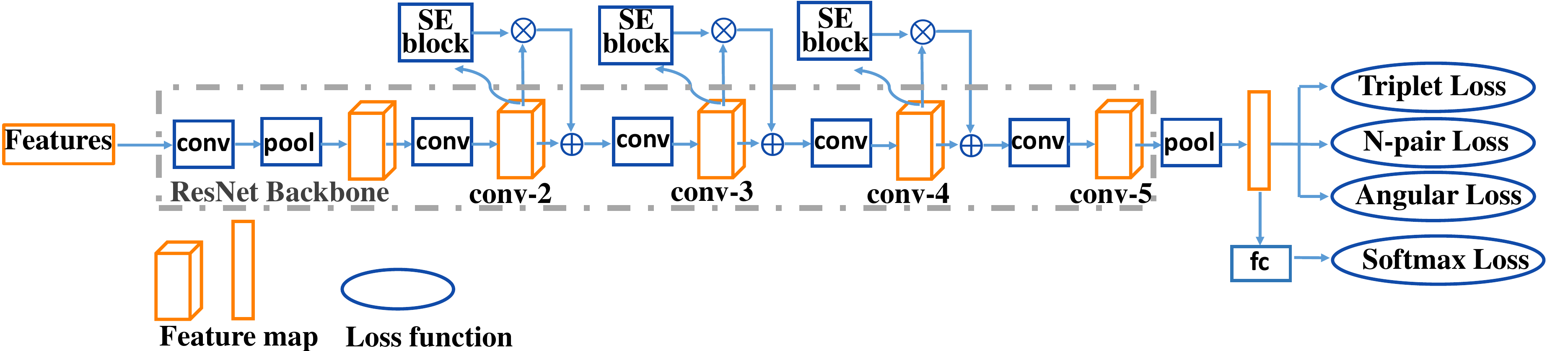}
\caption{The network architecture for training: its backbone network is ResNet-50; the pooling layers are all spatial average pooling; and the SE block is an attention module which is described in Figure~\ref{fig:se}; and has four loss functions, i.e., softmax loss, triplet loss, n-pair loss and angular loss.}
\label{fig:arc}
\end{figure*}

\section{Method}
\label{sec:mec}

In this section, we first describe the architecture of our network in Figure~\ref{fig:arc}. As the combination of ResNet \cite{he2016deep} and SENet \cite{zhang2016end, xie2019sparse} has been proven to achieve a good performance on the person re-identification (re-ID) task \cite{wang2018mancs}, we will embed the SE block in the ResNet to explore the channel-wise relationship for the SV task. Finally, we will analyze the previous metric losses and combine them to complement each other.

\subsection{Training Architecture}

As shown in Figure~\ref{fig:arc}, the training architecture of our method can be divided into two components: the ResNet-50 backbone and the loss functions. Here, the ResNet-50 backbone serves as a multi-scale feature extractor. In the second component, we separately calculate the triplet loss, N-pair loss, angular loss and softmax loss, and then devise a combination of those losses to optimize our network.

\subsection{SENet block}

In previous studies \cite{zhang2019seq2seq, zhou2019cnn}, the importance of attention has been proven in SV. The network will pay more attention to the discriminative local regions for SV. In recent works, the squeeze-and-excitation network (SENet) and Mancs \cite{wang2018mancs} illustrate that different channels of a feature map play different roles in specifying objects/parts. Taking those into consideration, we will introduce the SE block to the network in order to improve the performance of the network.

\begin{figure}[ht]
\centering
\includegraphics[width=1.0\columnwidth]{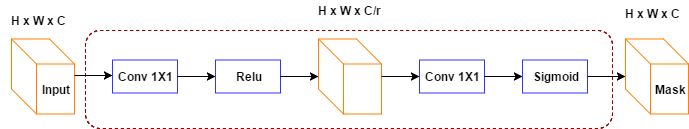}
\caption{The Squeeze-and-Excitation (SE) block in our SV network.}
\label{fig:se}
\end{figure}

As illustrated in Figure~\ref{fig:se}, the proposed SE block discards the pooling layer and replaces fully-connected layers with $1 \times 1$ convolutional layers to regain the spatial information \cite{wang2018mancs}. Given the input feature map $F_{i}$ of SE block, the output attention map $M$ can be computed as follow:
\begin{equation}
M = \text{Sigmoid} \left( \text{Conv}( \text{ReLU} (\text{Conv}(F_{i}))) \right),
\end{equation}
where the two Conv operators are $1 \times 1$ convolution. The roles of these two $1 \times 1$ convolutional layers are various. The inner one is used for squeeze and the outer one is used for excitation. The $1 \times 1$ convolution kernel can greatly increase the nonlinear characteristics (using the nonlinear activation function followed by the non-loss resolution) while keeping the feature map scale constant (i.e., without loss of resolution), making the network very deep). Via the SE block, we can obtain the feature map $F_{o}$ with attention information as
\begin{equation}
F_{o} = F_{i} \times M + F_{i}.
\end{equation}
With the attention feature maps added to the original feature map, it is believed that the discriminative information is emphasized. 

\subsection{Triplet loss}

At each training iteration, we sample a mini-batch of triplets, for each of which $T = (X_{a}, X_{p}, X_{n})$ consists of an anchor point $X_{a}$, associated with a pair composed of a positive sample $X_{p}$ and a negative sample $X_{n}$. The goal of triplet loss is to push away the negative point $X_{n}$ from the anchor $X_{a}$ by a distance margin $m > 0$ compared to the positive $X_{p}$. Triplet loss is usually defined as follows:
\begin{equation}
L_{tri}  = \left[ ||X_{a} - X_{p}||^2 + m - ||X_{a} - X_{n}||^2 \right]_+.
\label{eq:l_tri}
\end{equation}

Although triplet loss has achieved good results in many tasks, it has strict requirements on sampling strategies and takes a long training time to converge. Therefore, this paper introduces n-pair loss \cite{sohn2020distance} and angular loss \cite{wang2017deep}.

\begin{figure*}[htp]
\centering
\includegraphics[width=0.8\linewidth]{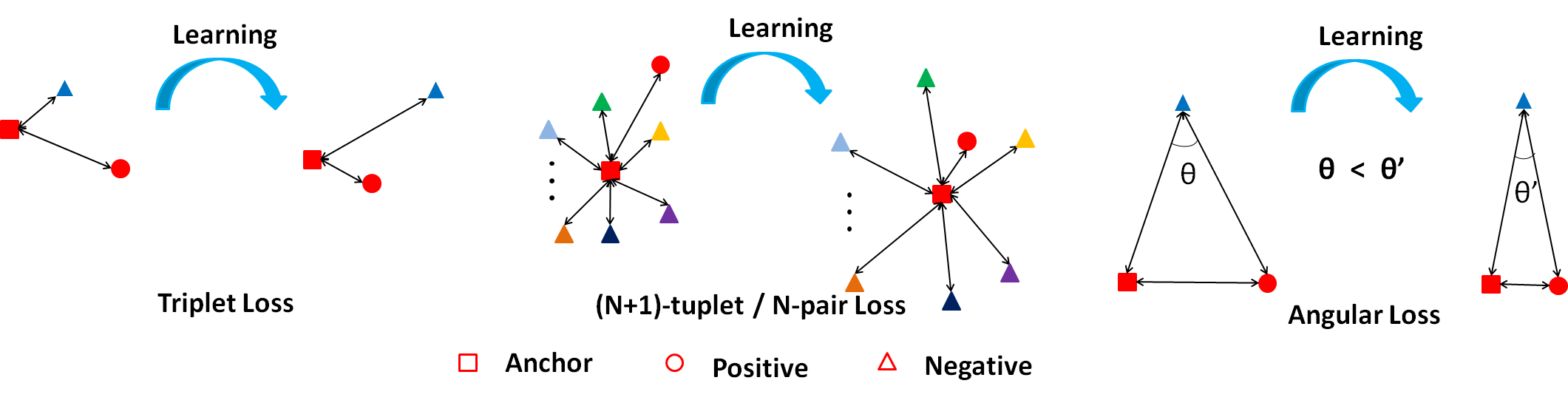}
\caption{Illustration of triplet loss, (N+1)-tuplet loss, n-pair loss and angular loss.}
\label{fig:loss}
\end{figure*}

\subsection{N-pair loss}

The traditional triplet loss only pays attention to the information of one negative sample in each optimization. Our assessment is that the training is slow and the information concerned is not comprehensive enough, so we introduce ($N+1$)-tuplet loss that optimizes the identification of a positive example from $N-1$ negative examples. By comparing with Figure~\ref{fig:loss}, we can observe that the traditional triplet loss is only a special case of ($N+1$)-tuplet loss, where $N=2$.
\begin{equation}
L_{(N+1)-\text{tuplet}} = \log(1+\sum_{i=1}^{N-1} \exp(f^{\top}f_i - f^{\top}f)),
\end{equation}
where $f$ is an embedding kernel defined by the deep neural network. To reduce the training burden while making full use of each batch of training samples, we propose a new training strategy. The corresponding (N+1)-tuplet loss, which we refer to as the n-pair loss, can be formulated as follows:
\begin{equation}
L_{\text{n-pair}} = \sum_{i=1}^{N}\log(1+\sum_{i \ne j} exp(f_{i}^{\top}f_{j}^+ - f_{i}^{\top}f_{i}^+)).
\end{equation}

Both the triplet loss and the n-pair loss only consider the distance between the anchor and the positive example and the anchor and the negative example; however, they do not consider the distance between the positive and the negative example, so the information is not comprehensive. Therefore, this paper introduces a more easily trained loss function, i.e., angular loss \cite{wang2017deep}.

\subsection{Angular loss}

Let us first imagine such an example, assuming three points $X_{a}$, $X_{p}$ and $X_{n}$, and these three points form a triangle $\triangle apn$, whose edges are denoted as $d_{ap} = X_{a} - X_{p}, d_{an} = X_{a} - X_{n}, d_{np} = X_{n} - X_{p}$, the traditional triplet loss can be seen as $d_{ap} + m < d_{an}$. We consider that the anchor and the positive example share the same label, so we can also optimize $d_{ap} + m < d_{pn}$. Within the triangle  $\triangle apn$, our goal is to find a solution that satisfies  $d_{ap}  < d_{pn}$ and $d_{ap} < d_{an}$, taking into account the fact that we can set a threshold that guarantees $\angle n$ $ \leq \angle \alpha $, in which $\alpha$ is our pre-set threshold. 

\begin{figure}[ht]
\centering
\includegraphics[width=1\linewidth]{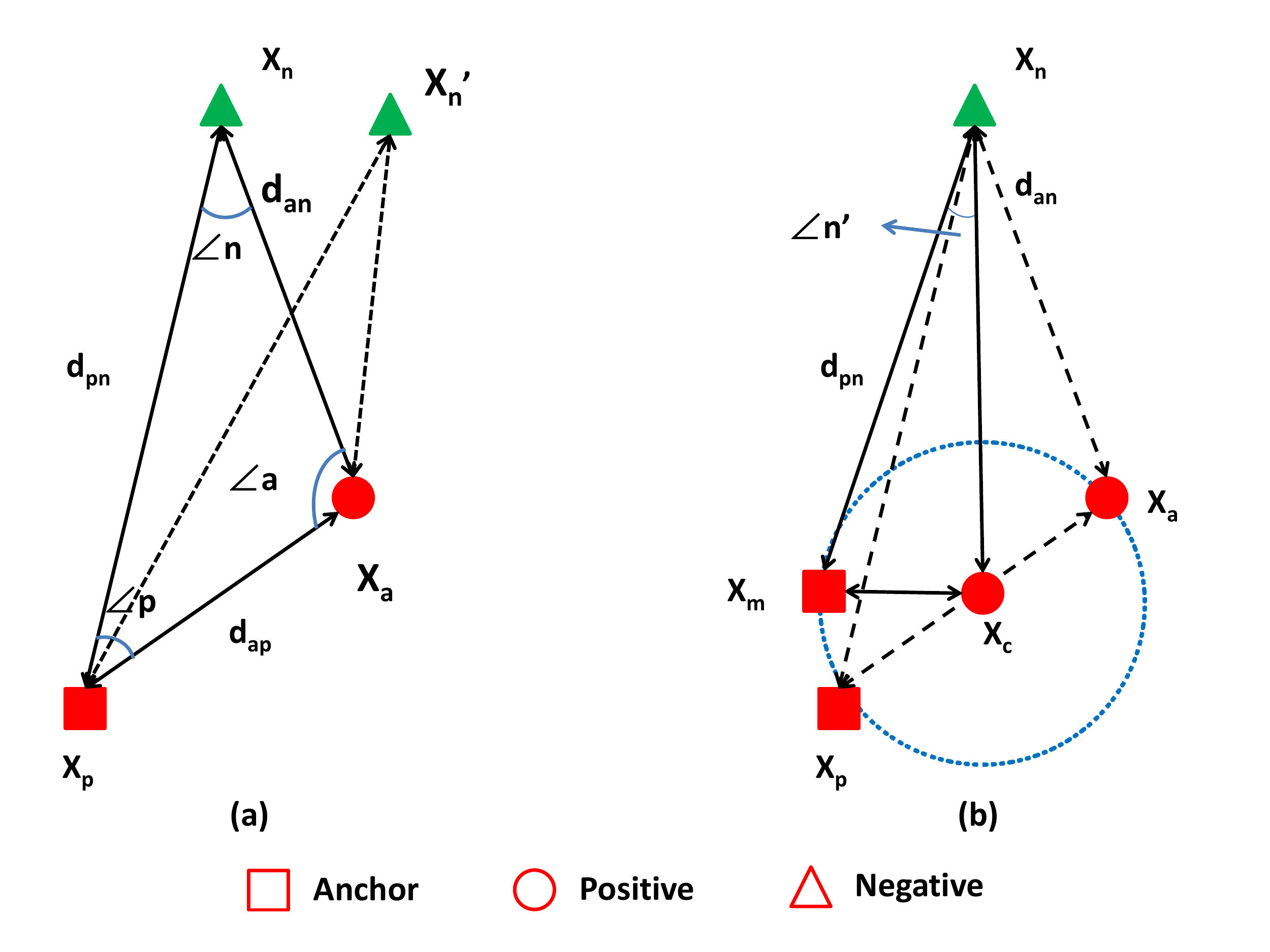}
\caption{Illustration of \( \triangle apn\) and \( \triangle cmn\) when computing angular loss.}
\label{fig:angular}
\end{figure}

However, in the actual optimization process, it is not very stable to consider only $\angle n$. Figure~\ref{fig:angular}~(a) is a special example, in $\triangle apn$, $\angle a > 90^{\circ}$, and $d_{an}$ does not decrease when $X_{n}$ is transformed to the position of $X_{n}'$. 

To fix this issue, we re-construct a triplet triangle for more stable optimization. Let us set point $X_{c}$ to the center point of $X_{a}$
and $X_{p}$, $X_{c} = (X_{a} + X_{p}) / 2$. $X_{m}$ is a point on the circle centered at $X_{c}$ and satisfies that $X_{m}X_{c}$ is vertical with $X_{c}X_{n}$. The triangle we re-construct is $ \triangle mcn$; at this point, the angle $ \angle n'$ is the angle we want to optimize. In the triangle $ \triangle mcn$, we can see that this formula is satisfied:
\begin{equation}
\sin \angle n' = \frac{||X_m - X_c||}{||X_n - X_c||} = \frac{||X_a - X_p||}{2||X_n - X_c||}
\leq \tan \alpha.
\end{equation}

Considering that $X_{c}$ is the center point of $X_{p}$ and $X_{a}$, therefore
$||X_m - X_c || =  || X_p - X_c ||  = 0.5 *  ||X_p - X_a || $. According to Eq.~\eqref{eq:l_tri}, we obtain the following equation:
\begin{equation}
{||X_p - X_a||}^2 \leq       {4||X_n - X_c||}^2 \tan ^2 \alpha.
\end{equation}

Inspired by the triplet loss our angular loss consists of minimizing the following hinge loss:
\begin{equation}
L_{ang} = max({4||X_n - X_c||}^2 \tan ^2 \alpha - {||X_p - X_a||}^2 \leq  ,0).
\end{equation}

\subsection{Multitask learning}

Multi-task learning has achieved good results in many areas, such as face verification \cite{lu2019experimental}, person re-ID \cite{ling2019improving, wang2018mancs}, SV \cite{monteiro2019combining}, metric learning \cite{lahoud20193d, xu2019multi}, etc. Considering that these tasks and the SV task have certain similarities, we can also use this method for training. The loss functions learned for metric learning in this paper have their advantages and disadvantages. The benefits of the combination of n-pair loss and angular loss have been verified in \cite{wang2017deep}. However, to the best of our knowledge, no studies exist that have used triplet loss with them. Our approach is to disclose some complementarity between them, which has been verified in our experiments. In addition, we use softmax loss with them to form a multi-task learning to further improve performance. As shown in Figure~\ref{fig:arc}, the two tasks share the same backbone network. In training, the corresponding three loss functions are optimized jointly. The overall loss is defined as follows:
\begin{equation}
L = \lambda_{n-pair}L_{n-pair} + \lambda_{tri}L_{tri} + \lambda_{ang}L_{ang} + \lambda_{soft}L_{soft},
\end{equation}
where $\lambda_{n-pair}$, $\lambda_{tri}$, $\lambda_{ang}$ and $\lambda_{soft}$  are weight factors for the loss functions.

\section{Experiments}
\label{sec:exp}

\subsection{Dataset}

We perform experiments on the VoxCeleb \cite{nagrani2017voxceleb} and VoxCeleb2 \cite{chung2018voxceleb2} datasets. We train our model on the training set of VoxCeleb2, which contains $1,128,246$ utterances from $5,994$ speakers. All models, including our model and the compared models, are tested on the testing set of VoxCeleb, which contains $37,720$ utterance pairs from $40$ speakers. The average duration of training and testing data are $8.24$s and $8.28$s, respectively. 

The utterances are extracted from videos on YouTube. Since these utterances originate from natural scenes, the signal quality is not very good and the background is noisy. Therefore, we firmly believe that if we can obtain good experimental results on this dataset, our method can be extended to more datasets and applied in the wild. 

\subsection{Data representation}

We first use traditional digital signal processing methods to characterize speech signals. The libROSA package \cite{mcfee2015librosa} is used for speech feature extraction. Spectrograms are generated in a sliding window fashion using a Hamming window with a width $20$ms and a step of $10$ms, in exactly the same manner as that of \cite{timmurphy.org}. Then we can obtain a vector whose dimension is the number of frames $\times 161$. Without loss of generality, we randomly intercept the three-second speech utterance, convert it to a $300 \times 161$ vector, and then copy it three times, constructing a $3 \times 300 \times 161$ vector similar to an image. If the duration of an utterance is less than $3$ seconds, we will copy the utterance to $3$ seconds.

\subsection{Training configurations}

We implement our network based on PyTorch~\cite{paszke2019pytorch}. {ResNet has been proven to be very effective in many tasks, so we also chose ResNet as the backbone. Considering that \cite{chung2018voxceleb2} chose ResNet-50 as the backbone, for fair comparison, we also chose the network as our backbone.} We take the ResNet-50 model pretrained on ImageNet as the backbone. We extract the conv-2, conv-3 and conv-4 feature maps to generate attention masks by SE blocks, and add them back into the mainstream. The last conv-5 feature map is used for generating the final utterance identity feature. We adopt the $PK$ sampling strategy to form every mini-batch. The values of both $P$ and $K$ are set to $64$ and $2$, respectively. This is to consider the set requirements of n-pair loss and angular loss and to ensure the difficulty of hard negative samples when training triple loss using a batch-hard sampling strategy. The activation function of the last convolutional layer is changed from ReLU to PReLU \cite{zuo2019dpgan}, which can effectively improve the over-fitting problem of the model. $\lambda_{n-pair}$, $\lambda_{soft}, \lambda_{tri}$ and $\lambda_{ang}$ are set to $0.5$, $0.1$, $1.0$ and $1.0$, respectively. The margin $\alpha$ in Eq.~\eqref{eq:l_tri} is set to $45^\circ$. We adopt the Adam optimizer with an initial learning rate of $3 \times 10^{-4}$ in our experiments to minimize the three losses. 

\subsection{Comparisons with the state-of-the-art methods}

We evaluated our proposed method against $13$ existing methods on VoxCeleb. As shown in Table~\ref{tab:main_results}, our model achieves the best result among the compared methods. As shown in the table there are currently two ways \cite{chung2018voxceleb2, okabe2018attentive} that can obtain an EER lower than $4\%$, and our method exceeds them.

\begin{table*}[htp]
  \caption{Speaker verification results on the standard VoxCeleb benchmark. Results of the compared methods are quoted from their original papers.}
  \vspace{2mm}
  \centering
    {
  \begin{tabular}{ccccc}
     \toprule
     \centering
     \normalfont \centering \bfseries Method & \centering \bfseries Dataset(s) & \centering \bfseries Architecture & \centering \bfseries Pooling  & \bfseries EER ($\%$) \\
     \midrule
     \centering
     VoxCeleb \cite{nagrani2017voxceleb} & \centering VoxCeleb1 & \centering PLDA+SVM & \centering Variable Length  & 8.80\\
     \centering
     VoxCeleb \cite{nagrani2017voxceleb} & \centering VoxCeleb1 & \centering VGG-M & \centering Variable Length  & 10.2\\
     \centering
     VoxCeleb \cite{nagrani2017voxceleb} & \centering VoxCeleb1 & \centering VGG-M & \centering Variable Length  & 7.80\\
     \centering
     CNN-TAP \cite{cai2018exploring} & \centering VoxCeleb1 & \centering Thin ResNet-34 & \centering Multi-Crop  & 5.27\\
     \centering
     i-vector \cite{okabe2018attentive} & \centering VoxCeleb1+PRISM & \centering PLDA & \centering Multi-Crop  & 5.39\\
     \centering
     TDNN \cite{okabe2018attentive} & \centering VoxCeleb1+PRISM & \centering TDNN & \centering Multi-Crop  & 4.70\\
     \centering
     LM \cite{okabe2018attentive} & \centering VoxCeleb1+PRISM & \centering TDNN & \centering Multi-Crop  & 4.69\\
     \centering
     LDE-ASoftmax \cite{cai2018exploring} & \centering VoxCeleb1 & \centering Thin ResNet-34 & \centering LDE  & 4.41\\
     \centering
     TDNN \cite{okabe2018attentive} & \centering VoxCeleb1 & \centering TDNN & \centering Attentive Stat.  & 3.85\\
     \centering
     VoxCeleb2 \cite{chung2018voxceleb2} & \centering VoxCeleb1+PRISM & \centering ResNet-50 & \centering Aariable Length  & 4.19\\
     \midrule
     \centering
     VoxCeleb2 \cite{chung2018voxceleb2} & \centering VoxCeleb2 & \centering ResNet-50 & \centering Multi-Crop  & 4.43\\
     \centering
     VoxCeleb2 \cite{chung2018voxceleb2} & \centering VoxCeleb2 & \centering ResNet-50 & \centering Average Dist.  & 3.95\\
     \centering
     \bfseries Ours & \centering VoxCeleb2 & \centering ResNet-50 & \centering Average Dist. &  \bfseries 3.48\\
     \bottomrule
  \end{tabular}
  }
  \label{tab:main_results}
\end{table*}

\subsection{Ablation study}

We further perform several ablative experiments to verify the effectiveness of each individual loss function of our proposed model. The results are shown in Table~\ref{tab:ablation_losses}. We always apply softmax classification loss, which is simple and can stabilize the training process. The EER of triplet loss, N-pair loss and angular loss are $5.00\%$, $4.90\%$ and $5.72\%$, respectively. 
We found that angular loss can not achieve better results than triplet loss, which occurs triplet loss employs a semi-hard sampling strategy while the angular loss function does not. Integrating triplet loss, N-pair loss and angular loss can obtain the best results, which is $3.48\%$ EER.

\begin{table}[htp]
    \centering
    \caption{Ablation studies of the loss functions.}
      \vspace{2mm}
    \centering
    \begin{tabular}{cccccc}
    \toprule
      Triplet loss & \checkmark  &  & & \checkmark \\
      N-pair loss &  & \checkmark & &  \checkmark \\
      Argular loss &  &  & \checkmark &  \checkmark \\
      Softmax loss & \checkmark & \checkmark & \checkmark & \checkmark \\
      \midrule
      EER ($\%$) &  $5.00$  &  $4.90$  &  $5.72$  & \bfseries $3.48$ \\
      \bottomrule
    \end{tabular}
    \label{tab:ablation_losses}
\end{table}

Through the experiments, we can see that the best results are obtained by training multiple losses together. We analyze this combination approach because there is some complementarity between the three losses. First, if a strict sampling strategy and a large amount of training time are adopted, triplet loss can achieve good results, but the selection of margin is difficult during training. The sampling strategy we use in this paper is the semi-hard example mining method which was presented in \cite{schroff2015facenet}. In addition, triplet loss compares one positive sample with one negative sample during training, and we contend that the compared examples in each batch are insufficient.
Second, the n-pair loss can make full use of the training data in a batch by considering the pairwise information between negative samples, which speeds up the training process and achieves good results but it does not require a good sampling strategy during training.
Third, compared with the traditional triple loss, the n-pair loss is defined on the absolute distance between points. The proposed angular constraint offers three advantages: $1)$ Angle is a similarity-transform invariant metric, which is insensitive to the magnitude of features. $2)$ The original triple loss only considers the two sides of the triangle but $\angle n$ takes into account the three sides of the triangle, so the information considered is more comprehensive. $3)$ In the original triplet loss, it is difficult to set a standard threshold $\alpha$.However, in our angular loss, setting the threshold $\alpha$ is relatively easy because it has a clear geometric meaning. However, the disadvantage of angular loss and n-pair loss is the same, that is, there is no good sampling strategy, which leads to the emergence of many invalid training data during the training. Finally, based on the above analysis, we believe that training the three loss functions together can complement each other, which was also verified in the experiment.

In addition, through the comparison experiments, we found that adding the SENet block can achieve a clear performance improvement and the EER decreases from $3.78\%$  to $3.48\%$. {The role of the SE block is that it can significantly improve the discrimination of local features, which has also been confirmed in other tasks \cite{li2020classification, yan2020deep, li2020classification}}.

\subsection{Training and Testing Time}

All experiments are run on a server with two TITAN XP GPUs. We find that it takes approximately $9$ days to train in the VoxCeleb2 dataset using all the loss functions. To reduce the training time, our strategy is to use softmax pre-training to initialize the weights of the network which takes approximately $2$ days, and then fine-tune the network which takes approximately $5$ days. Overall, this saves $2$ days of training time and obtains similar results to the models trained for $9$ days. In the testing stage, we take the same settings as in \cite{chung2018voxceleb2} and sample $10$ three-second temporal crops from each test segment, compute the distances between every possible pair of crops (10 $\times$ 10 = 100) from the two speech segments, and use the mean of the 100 distances which takes approximately $5$ hours.

\section{Conclusion}
\label{sec:con}

In this paper, we investigate the application of triplet loss, n-pair loss and angular loss in the task of text-independent speaker verification (SV) based on deep learning. A single-metric loss function often has certain limitations and we found that multiple metric losses often have certain complementarity. Therefore, better results can be achieved by combining multiple metric learning losses. To the best of our knowledge, this is the first time that multiple metric learning loss functions have been applied to the field of text-independent SV. Inspired by the fact that attention can localize the most discriminative local regions for SV, we introduce the SE block into the network to further improve performance. Large-scale experiments show that our method has achieved very competitive results on the VoxCeleb1 test set and ablative studies confirm the effectiveness of the proposed deep multi-metric learning. To facilitate the research of text-independent SV, we have release all the training and testing source codes as well as pretrained models. We believe that our method can also be applied to tasks such as image retrieval, face recognition, and person re-identification, which will be explored our in future work.

\section*{Acknowledgments}
This work was supported by NSFC (No. 61876212, No. 61733007, No. 61773176 and No. 61572207) and National Key R\&D Program of China (No. 2018YFB1402600). We sincerely thank the anonymous reviewers for their helpful reviews. 

\section*{References}
\bibliographystyle{apa}
\bibliography{TiSV}

\end{document}